\documentclass{elsart}

\usepackage{graphicx}

\usepackage{amsmath}

\begin{document}

\begin{frontmatter}



\title{Plug Conveying in a Vertical Tube}

\author[ICP]{Martin Strau\ss}
\author[ICP]{Sean McNamara}
\author[ICP]{Hans J.\ Herrmann}
\author[MAK]{Gerhard Niederreiter}
\author[MAK]{Karl Sommer}

\address[ICP]{Institut f\"ur Computerphysik, Universit\"at Stuttgart,
70569 Stuttgart, GERMANY}
\address[MAK]{Lehrstuhl f\"ur Maschinen- und Apparatekunde, TU M\"unchen,
85350 Freising, GERMANY}

\begin{abstract}
Plug conveying along a vertical tube has been investigated through simulation,
using a discrete element simulation approach for the granulate particles
and a pressure field approach for the gas. The result is compared with
an experiment. The dynamics of a plug are described by porosity,
velocity and force profiles. Their dependence on simulation parameters
provides an overall picture of plug conveying.
\end{abstract}

\begin{keyword}
plug conveying \sep dense phase \sep pneumatic transport \sep granular medium

\PACS 47.55.Kf \sep 45.70.Mg
\end{keyword}
\end{frontmatter}

\newcommand{\fd}[2]{\frac{d {#1}}{d {#2}}}
\newcommand{\fdd}[2]{\frac{d^2 {#1}}{d {#2}^2}}
\newcommand{\pd}[2]{\frac{\partial {#1}}{\partial {#2}}}
\newcommand{\pdd}[2]{\frac{\partial^2 {#1}}{\partial {#2}^2}}
\newcommand{\grad}{\vec{\nabla}}
\renewcommand{\div}{\vec{\nabla}}
%
%
\section{Introduction}
A quite common method for the transportation of 
granular media is pneumatic conveying, where grains are driven
through pipes by air flow.
Practical applications for pneumatic conveying can be found in food industry
and in civil and chemical engineering.
One distinguishes two modes of pneumatic conveying: dilute and dense phase
conveying.
Dilute phase conveying has been studied in much detail%
~\cite{ALA0126,DAS0130,RAU0121,VAN0062,MAS0056,BIL0153,YAM0039,MAS0138}
and is well understood.
The grains are dispersed and dragged individually by
the gas flow and the interaction between grains is small.
This is not true for the dense phase conveying,
where the particle interaction is important
and where particle density waves can be observed.
Two modes of dense phase conveying can be distinguished,
depending on the orientation of the pipe transporting the grains.
In the horizontal transport the granular medium separates into two
layers, the moving bed at the bottom of the pipe and the irregular
transport by traveling dunes or slugs%
~\cite{TSU0093,TOM0147,MAS0154,ZHU0120,VAS0136}.
In the vertical transport the granular medium
forms strands at medium gas velocities%
~\cite{WAS0057,TSU0058,HEL0059,AGR0063,TSU0091}
and plugs at low gas velocities with high mass loads%
~\cite{KAW0090,JAW0104,ZHU0120}%
; this mode is called plug conveying.
Currently plug conveying is gaining importance in industry, because
it causes a lower product degradation 
and pipeline erosion than dilute phase conveying.

Unfortunately, current models~\cite{KON0095,SIE0096}
of plug conveying disagree
even on the prediction of such basic quantities
as the pressure drop and the total mass flow, and
these quantities have a great impact in the industrial application.
One of the reasons for the lack of valid models is that
it is difficult to study plugs experimentally in a detailed way.
Usually experimental setups are limited to the measurement of the
local pressure drop, the total mass flux and the velocity of plugs.
Most promising experimental studies have been performed by electrical capacity
tomography~\cite{JAW0104,ZHU0120} and 
stress detectors~\cite{NIE0111,NIE0112,VAS0136}.
Simulational studies are handicapped by the high computational cost
for solving the gas flow and the particle-particle interaction,
and are therefore mostly limited to two dimensions.
For the dense phase regime simulations have been done for 
bubbling fluidized beds~\cite{YEM0122,TSU0092,KAW0089,HUI0061,GOL0125,YUU0159}, which show for high gas velocities
first signs of pneumatic transport~\cite{LIM0115,XUB0118,HOO0119},
for the strand type of conveying~\cite{HEL0059,TSU0058,TSU0091,YON0098},
and slugs in horizontal transport~\cite{TSU0093,LEV0155}.
Simulations of vertical transport which show plugs with considerable length
are rare.
There is only one simulation of a short pipe in two dimensions 
by Tsuji et al.~\cite{KAW0090}
and a picture of a single plug from a simulation with 50 particles 
in three dimensions by Ichiki et al.~\cite{ICH0162}.

The goal of this paper is to provide a detailed view of plugs, by using
a discrete element simulation combined with a solver for the pressure drop.
This approach provides access to important parameters like the porosity
and velocity of the granulate and the shear stress on the wall
at relatively low computational cost.
Contrary to the experiments,
it is possible to access these parameters 
at high spatial resolution and without influencing the process of 
transportation at all. 
Additionally to plug profiles, characteristic curves of the pressure drop 
and the influence of simulation parameters are measured.
The simulation results are compared with and verified by experiments.
%
%
%
%
\section{Simulation Model}
Plug conveying is a special case of the two phase flow of grains and
gas. It is therefore necessary to calculate the motion of both phases,
as well as the interaction between them.
In the following, we explain how our algorithm treats each of these problems.
\subsection{Gas Algorithm}
The model for the gas simulation was first introduced by
McNamara and Flekk{\o}y \cite{MCN0003} and has been implemented for
the two dimensional case
to simulate the rising of bubbles within a fluidized bed.
For the simulation of plug conveying we developed a three dimensional version
of this algorithm.

The algorithm is based on the mass conservation of the gas
and the granular medium.
Conservation of grains implies that
the density $\rho_p$ of the granular medium obeys
\begin{equation}
\pd{\rho_p}{t}+\vec{\nabla}\cdot(\vec{u}\rho_p)=0,\qquad \rho_p=\rho_{s}(1-\phi),
\end{equation}
where the specific density of the particle material is $\rho_{s}$,
the porosity of the medium is $\phi$
(i.e. the fraction of the space available to the gas),
and the velocity of the granulate is $\vec{u}$.

The mass conservation equation for the gas is
\begin{equation}
\pd{\rho_g}{t}+\vec{\nabla}\cdot(\vec{v}_g\rho_g)=0,
\qquad \rho_g\propto \phi P,\quad\qquad 
\end{equation}
where $\rho_g$ is the density of the gas and $\vec{v}_g$ its velocity.
This equation
can be transformed into a differential equation for the gas pressure $P$
using the ideal gas equation, together with the assumption of uniform
temperature.

The velocity $\vec{v}_g$ of the gas is related to
the granulate velocity $\vec{u}$ through the d'Arcy relation:
\begin{equation}
-\vec{\nabla}P=\frac{\eta}{\kappa(\phi)}\phi(\vec{v}_g-\vec{u}),
\end{equation}
where $\eta$ is the dynamic viscosity of the air and $\kappa$ is the
permeability of the granular medium. This relation was first given
by d'Arcy in 1856 \cite{DAR0094}.
For the permeability $\kappa$ the Carman-Kozeny relation \cite{CAR0084}
was chosen,
which provides a relation between the porosity $\phi$, the particle diameter $d$
and the permeability of a granular medium of monodisperse spheres,
\begin{equation}
\kappa(\phi)=\frac{d^2\phi^3}{180(1-\phi)^2}.
\end{equation}
After linearizing around
the normal atmospheric pressure $P_0$ the resulting differential equation
only depends on the relative pressure $P^\prime$ ($P=P_0+P^\prime$),
the porosity $\phi$ and the granular velocity $\vec{u}$,
which can by derived from the particle simulation,
and some constants like the viscosity~$\eta$:
\begin{equation}\label{eqn:dgl}
\pd{P^\prime}{t}=\frac{P_0}{\eta\phi}\vec{\nabla}(\kappa(\phi)\vec{\nabla}P^\prime)-\frac{P_0}{\phi}\vec{\nabla}\vec{u}.
\end{equation}
This differential equation can be interpreted as a diffusion equation with
a diffusion constant $D=\phi\kappa(\phi)/\eta$.
The equation is solved numerically, using a
Crank-Nickelson approach for the discretization. Each dimension is integrated
separately.

The boundary conditions are imposed by
adding a term $\mp S$ on the right hand side of equation (\ref{eqn:dgl})
at the top and the bottom of the tube, where $S\propto v_gP_0$.
This resembles a constant gas flux with velocity $v_g$
at a pressure $P_0$ into and out of the tube.
\subsection{Granulate Algorithm}
The model for the granular medium simulates each grain
individually using a discrete element simulation (DES).
For the implementation of the discrete element simulation we used 
a version of the molecular dynamics method described by 
Cundall~\cite{CUN0172}.
The particles are approximated as monodisperse spheres, rotations in
three dimensions are taken into account.

The equation of motion for an individual particle is
\begin{equation}
m\ddot{\vec{x}}=m\vec{g}+\vec{F}_{c}-\frac{\grad P}{\rho_s(1-\phi)},
\end{equation}
where $m$ is the mass of a particle, $\vec{g}$ the gravitation constant and
$\vec{F}_{c}$ the sum over all contact forces. The last term,
the drag force, is assumed to be a volume force
given by the pressure drop $\grad P$ and the local mass density of
the granular medium $\rho_s(1-\phi)$.

The interaction between two particles in contact is given
by two force components:
a normal and a tangential component with respect to the particle surface.
The normal force is the sum of a repulsive elastic force (Hooke's law)
and a viscous damping.
The tangential force is proportional to the normal force
(sliding Coulomb friction) or a viscous damping.
The viscous damping is used when the relative
surface velocity of the particles in contact is small.
The same force laws are considered for the interaction between particles 
and the tube wall.
\subsection{Gas-Grain Interaction}
The simulation method uses both a continuum and a
discrete element approach. While the gas algorithm
uses fields, which are discretized on a cubic grid,
the granulate algorithm describes particles in a continuum.
A mapping is needed for the algorithms to interact.
For the mapping a tent function $F(\vec{x})$ is used:
\begin{equation}
F(\vec{x})=f(x)f(y)f(z),\qquad f(x)=\begin{cases}
  1-|x/l|, &  |x/l| \le 1, \\
  0, &  1<|x/l|,
\end{cases}
\end{equation}
where $l$ is the grid constant used for the discretization
of the gas simulation.

For the gas algorithm the porosity $\phi_j$ 
and the granular velocity $\vec{u}_j$
must be derived from the particle positions $x_i$ and velocities $v_i$,
where $i$ is the index of particle and
$j$ is the index for the grid node.
The tent function distributes the particle properties around
the particle position smoothly on the grid:
\begin{equation}
\phi_j=1-\sum_i F(\vec{x}_i-\vec{x}_j), \qquad
\vec{u}_j=\frac{1}{1-\phi_j}\sum_i \vec{v}_i F(\vec{x}_i-\vec{x}_j),
\end{equation}
where $x_j$ is the position of the grid point 
and the sum is taken over all particles.

For the computation of the drag force on a particle
the pressure drop $\grad P_i$ and the
porosity $\phi_i$ at the position of the particle are needed.
These can be obtained by a linear interpolation of the fields $\grad P_j$
and $\phi_j$ from the gas algorithm:
\begin{equation}
\phi_i=\sum_j \phi_j F(\vec{x}_j-\vec{x}_i), \qquad 
\grad P_i=\sum_j \grad P_j F(\vec{x}_j-\vec{x}_i),
\end{equation}
where the sum is taken over all grid points.
\clearpage
%
%
%
%
\section{Experimental Results}
\begin{figure}[h]
\includegraphics[width=3cm]{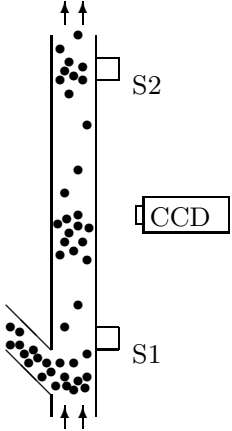}
\caption{\label{fig:experiment-setup}%
Sketch of the experimental setup.
Gas is inserted at a constant mass rate at the bottom of the tube,
the granulate enters from the left. S1 and S2 denotes pressure sensors.
A CCD camera is used to record the transport at mid-height.
}
\end{figure}
In the experiment we study the vertical pneumatic transport of wax beads of
diameter $d=1.41mm$, density $\rho_s=937kg/m^3$ and a Coulomb coefficient of 0.21.
The experimental transport
channel is a vertical tube (PMMA) of length $l=1.01m$ 
and of internal diameter $D_t=7mm$.
These parameters were chosen to be able to compare to simulations
with little computational effort possible.
The air is injected at a constant mass flow rate at the bottom of the tube.
The granular medium is injected from the side from a tube with a slope of $45^\circ$.

The wax beads, when leaving the vertical tube at the top,
are gathered and weighted. From this data the mass flow rate 
of the granular medium can be obtained.
The total pressure drop is detected through two pressure sensors,
one at the bottom (S1) and one at the top (S2).
The velocity and length of the plugs can be obtained by a CCD-camera 
(30 frames per second) positioned at the middle of the tube.

\begin{figure}[h]
\begin{center}
\includegraphics[width=10cm]{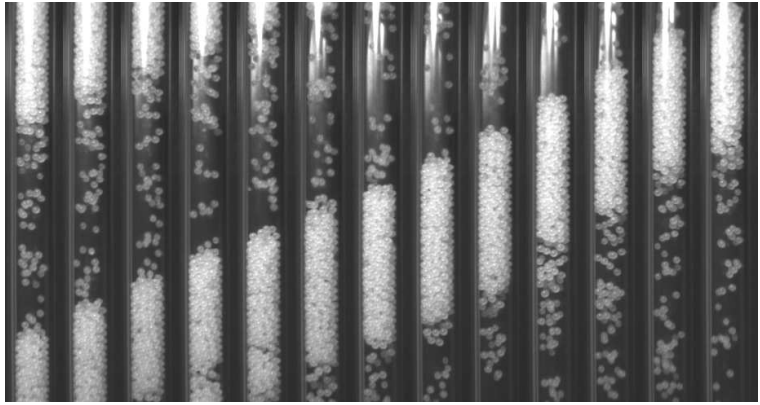}
\end{center}
\caption{\label{fig:expplugs}%
A series of photos showing a plug moving upwards.
The height of the shown tube is $9\,cm$, the frame~rate is $30\,Hz$,
from measurement series B.
}
\end{figure}

\begin{table}[h]
\begin{tabular}{ll|lll}
series & & A & B & C \\\hline
air flow &($l/min$) & $2.0$ & $2.2$ & $2.4$ \\ 
pressure drop &($hPa/m$)& $42\pm 2$ & $45\pm 2$ & $44.5\pm 2$ \\
granulate flow &($kg/h$)& $1.7\pm  0.15$ & $2.49\pm 0.02$ & $2.88\pm 0.08$\\
plug velocity &($m/s$)& 0.09-0.17 & 0.12-0.19 & 0.18-0.33 \\
plug length &($cm$)& 0.6-4 & 1.7-9 & 0.5-4 
\end{tabular}
\caption{\label{tab:experiment}%
Experimental results for vertical plug conveying in a tube of length $l=1.01\,m$
and diameter $D_t=7\,mm$ using wax beads ($d=1.4\,mm$) as granular medium.
}
\end{table}
Plug conveying is observed, as shown in figure~\ref{fig:expplugs}.
The range of parameters for plug conveying in the experiment
were rather limited. The plugs often get stuck in the tube, interrupting
the plug transport completely. For long runs and repeated use of grains,
electrostatic forces influenced the results.
Measurement series for three different air flow rates were made.
The results for the total pressure drop and the granular flow are given
in table \ref{tab:experiment}, for more details see \cite{NIE0173}.

\clearpage
%
%
%
%
\section{Simulational Results}
The setup for the simulation is almost the same as for the experiment.
The main difference is:
particles are injected from the bottom at a constant mass flow rate. 
As default the mass flow rate of the granular medium is chosen equal to 
the measured flow rate in the experiment ($2.49\,kg/h$).
Default parameters for the particles are:
diameter $d=1.4\,mm$, density $\rho_s=937\,kg/m^3$,
Coulomb coefficient $\mu=0.5$ and restitution coefficient $e=0.5$.
Simulations are preformed for half tube length $L_t=0.525\,m$ and 
tube diameter $D_t=7\,mm$.
The resulting gas volume has been discretized into 150x2x2 grid nodes,
which corresponds to a grid constant of $3.5\,cm$.
The gas pressure is set to $P_0=1013.25\,hPa$, Simulations are preformed
for gas viscosities $\eta$ from $0.045\,cP$ to $0.085\,cP$
and gas flows $\dot{V}$ between $1.1\,l/min$ and $9.2\,l/min$.
The gas flow is usually given as the 
superficial gas velocity~$v_s=\phi v_g=4\dot{V}/\pi D_t^2$~\cite{HON0128},
which is the equivalent gas velocity for an empty tube.

\begin{figure}[h]
\includegraphics[width=14cm]{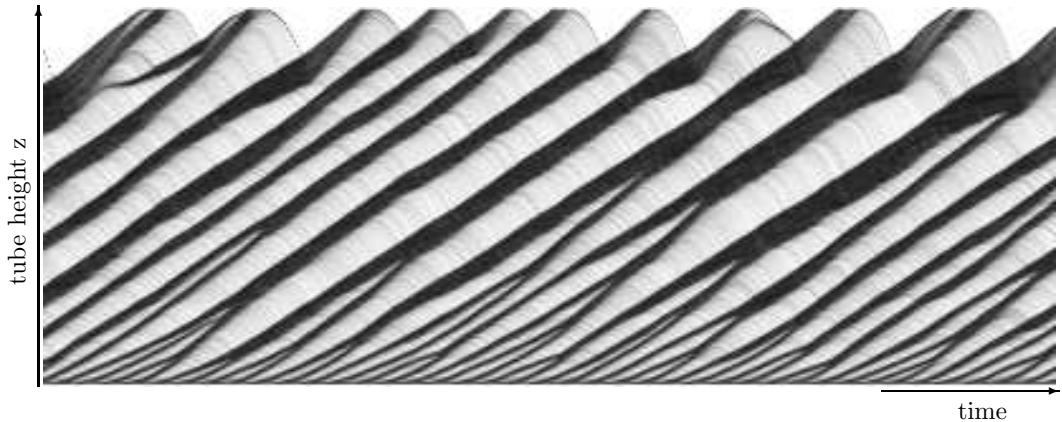}
\caption{\label{fig:rho-W0126}%
Spatio-temporal image of the porosity along the vertical tube.
The dark regions correspond to low porosity, light regions to high
porosity.
The entire tube ($52.5\,cm$) is displayed, the elapsed time is~$4\,s$.
Default parameters for the simulation were used, the superficial gas velocity
is $1\,m/s$, the gas viscosity is $0.0673\,cP$,
the Coulomb coefficient is $0.5$.
Particles are introduced at the bottom and on average 3600 particles
are within the tube at a pressure drop of $39\,hPa/m$.
}
\end{figure}

The flow in the experiment is turbulent (particle $Re\approx 65$).
In the simulation, an effective gas viscosity is used to account
for the turbulence.
For an effective gas viscosity $\eta=0.0673\,cP$, 
a gas flow $\dot{V}=2.3\,l/min$
and a Coulomb coefficient $\mu=0.5$
plug conveying is observed as shown in figure~\ref{fig:rho-W0126}.
The observed pressure drop $39\,hPa/m$, the plug velocity $0.24\,m/s$ and 
the plug length 2-4$\,cm$ fit well with the experimental results.

As pointed out in the introduction, only very limited simulations of 
vertical plug conveying have been published.
Vertical transport has been studied by Tsuji~\cite{KAW0090}, but his
simulation was limited to a short pipe in two dimensions ($L_t\approx 3d$).
In his case the tube contained about 300 particles,
which is in our case about the number of particles within one plug.
Three dimensional plugs of about 50 particles have also been 
found by Ichiki~\cite{ICH0162}, but no analysis of them was done.
Thus our simulations are the first to show full featured vertical
plug conveying at the level of grains in three dimensions.

\subsection{Characteristic curves}
\begin{figure}[h]
\begin{center}
\includegraphics[width=10cm]{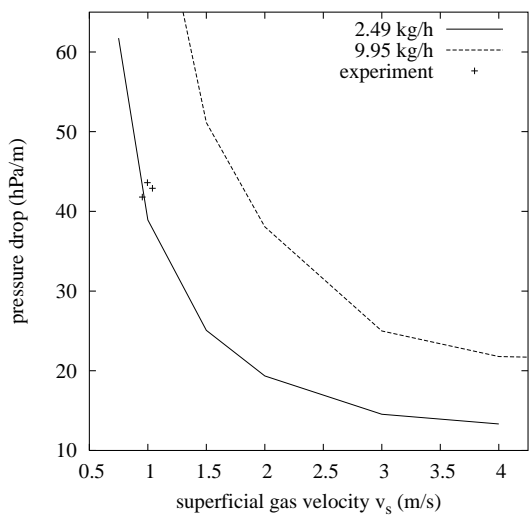}
\end{center}
\caption{\label{fig:v-dp}%
Total pressure drop against superficial gas velocity for different
granular mass flows. Plotted are characteristic curves
from the simulation for the granular mass flows $2.49\,kg/h$ and $9.95\,kg/h$,
and the three data points from the experiment with granular mass flows $1.7-2.9\,kg/h$.
For mass flow $2.49\,kg/h$ plugs stick when the superficial gas velocity
is below $0.75\,m/s$.
}
\end{figure}

The ``characteristic curves'' of a pneumatic transport system
are plots of the pressure drop against the superficial gas velocity
$v_s=\phi v_g$ for different mass flows of the granulate.
This kind of diagram is
highly dependent on the material characteristics of the tube wall and
the granulate and can be used to predict the overall transport performance
for given parameter sets.
Such a diagram, from data from our simulation,
is shown in figure~\ref{fig:v-dp}.

The diagram provides the typical qualitative behavior for pneumatic transport.
From top left to bottom right with increasing superficial gas velocity, four regions
can be distinguished.
First, for small superficial gas velocities~($v_s<0.5\,m/s$),
there is bulk transport.
The tube is completely filled with granulate, so the pressure drop is high.
Nevertheless the drag force on the bulk is 
too small to support the weight of the granular medium. 
In this case the transport comes through the enforced granular mass flow at the bottom.
\begin{figure}[h]
\includegraphics[width=14cm]{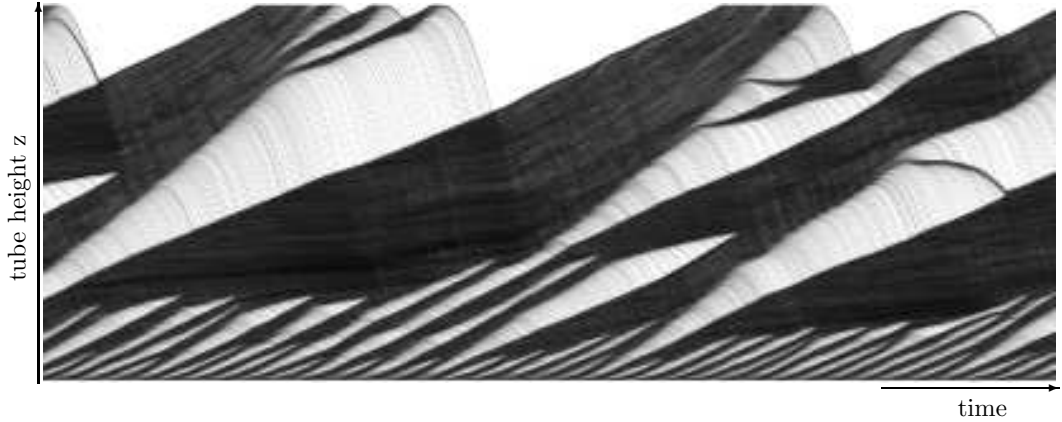}
\caption{\label{fig:rho-W0134}%
Spatio-temporal image of the porosity along the vertical tube.
The dark regions correspond to low porosity, light regions to high
porosity.
The entire tube ($52.5\,cm$) is displayed, the elapsed time is~$4\,s$.
A lower superficial gas velocity ($0.75\,m/s$)
has been used than in figure~\ref{fig:rho-W0126}.
On average 4900 particles are within the tube at a pressure drop of $62\,hPa/m$.
Due to the low drag force the plugs quickly coalesce into very long plugs.
}
\end{figure}

For slightly higher gas velocities~($0.5\,m/s\le v_s<0.75\,m/s$)
the first plugs appear.
These plugs are often not able to compensate the force from material
falling onto them.
They usually collapse into big plugs sticking in the tube or 
their motion is reduced considerably (fig.~\ref{fig:rho-W0134}).
In this region the pressure drop decreases 
rapidly with increasing gas velocity.

For moderate velocities plug conveying is observed $0.75\,m/s\le v_s<4\,m/s$
(fig.~\ref{fig:rho-W0126}).
The qualitative behavior of the granular medium in this regime is the same in the
experiment and the simulation.
The particles injected at the bottom organize into plugs.
After a short acceleration at the bottom plugs move upwards
with a constant velocity.
When two plugs collide, they combine into a single plug.
A plug always looses particles at the bottom
and usually maintains its length by collecting particles at the top.
A plug disintegrates when it gets too small.
The porosity of the granular medium within a plug
is close to the minimum porosity, the edges are sharp and the space between
plugs is rather empty ($\phi>90\%$).
As one can see in figure~\ref{fig:v-dp}
the experimental data were measured for a superficial gas velocity
at the lower limit for plug conveying.
In this region the pressure drop is highly dependent on 
the material parameters ($\eta$,$P_0$,$\mu$).

\begin{figure}[h]
\includegraphics[width=14cm]{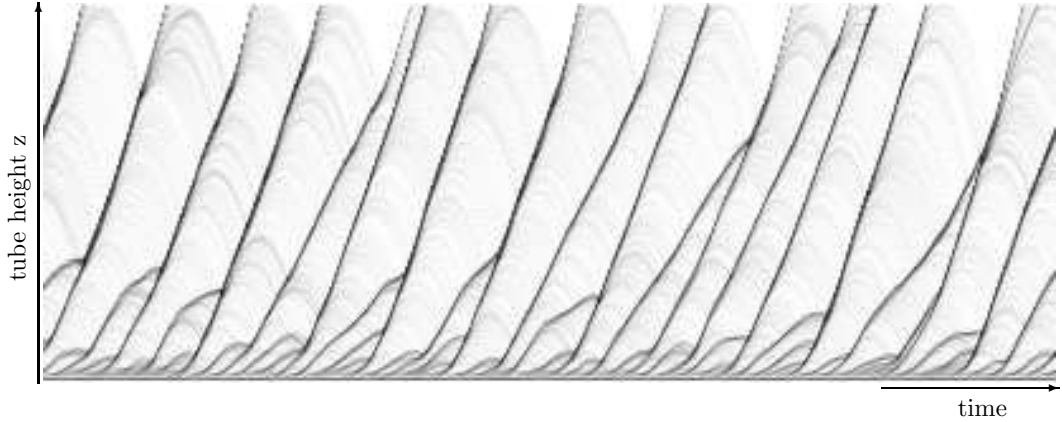}
\caption{\label{fig:rho-W0136}%
Spatio-temporal image of the porosity along the vertical tube.
The dark regions correspond to low porosity, light regions to high
porosity.
The entire tube ($52.5\,cm$) is displayed, the elapsed time is~$4\,s$.
A higher superficial gas velocity (4\,m/s)
has been used compared to the spatio-temporal image in figure~\ref{fig:rho-W0126}
($1\,m/s$).
On average 850 particles are within the tube at a pressure drop of $13\,hPa/m$.
}
\end{figure}

For high superficial gas velocities~($v_s>4\,m/s$) the tube is almost empty
(fig.~\ref{fig:rho-W0136}),
in the simulation the particles are pushed out as small plugs.
Their porosity increases with the superficial gas velocity.
In this region the simulation method underestimates the pressure drop,
because it does not consider the increasing drag force on single particles,
which in the experiment dominates in this region.
The regions as described above shift with changes of the simulation parameters.

A nearly proportional relation is observed
between the total pressure drop and the total number of particles in the tube.
This can be explained by the observation that
the amount of particles between the plugs is small and
most particles are densely packed at a well 
defined porosity in the plugs.
Through the d'Arcy's law,
the total pressure drop depends linearly on the tube length
filled with this porosity. The pressure drop on the granulate between the
plugs is negligible and causes only a small deviation from the proportional relation.
At high gas velocities this is no longer true, because there is no dominating
porosity for the plugs.

In the following the dependence of the pressure drop on 
the air viscosity $\eta$, the atmospheric pressure $P_0$ and the
Coulomb coefficient $\mu$ is discussed.
For the parameter studies the superficial gas velocity has been
fixed to $1\,m/s$. This resembles the gas velocities used for
the experiment. For higher velocities the parameter dependence decreases.

The gas algorithm can be influenced by changing the atmospheric pressure $P_0$
or the diffusion constant $D$.
An increase in background pressure $P_0$ combined with a proportional increase
in superficial velocity leaves the pressure drop unchanged.
This can be deduced directly from equation (\ref{eqn:dgl})
by noting that both terms on the right hand side are proportional to $P_0$.

\begin{figure}[h]
\begin{center}
\includegraphics[width=10cm]{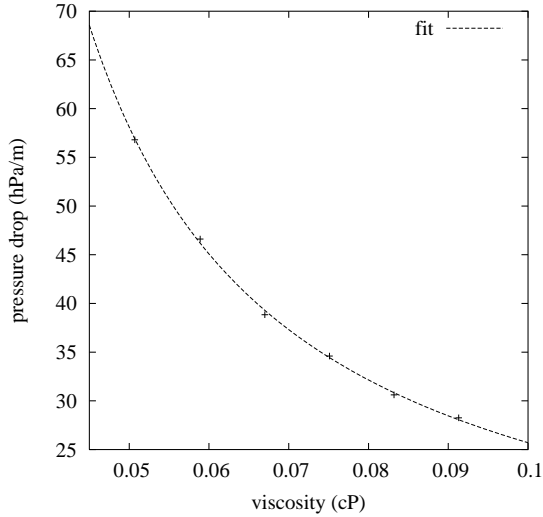}
\end{center}
\caption{\label{fig:vis-dp}%
Dependence of the pressure drop on the dynamic viscosity $\eta$
of the gas at superficial gas velocity $1\,m/s$.
The data points can be fitted by a hyperbola.
}
\end{figure}

The diffusion constant $D\propto d^2/\eta$ can be changed
through the particle diameter $d$ and the viscosity $\eta$.
Therefore it is sufficient to analyze the parameter space for the viscosity
at a constant diameter as shown in figure~\ref{fig:vis-dp}.

The viscosity dependence of the pressure drop can be described by a hyperbola.
The pressure drop converges for high viscosities to a finite value.
In this case the pressure drop supports the weight of the granulate
embedded in the fluid and depends only on the velocity of the gas
and the enforced mass flow of the granular medium. 
For small viscosities plug conveying starts to break down to bulk transport.
The pressure drop diverges at the viscosity where the drag force
can no longer balance the weight and friction of the bulk.

The parameter of the particle simulation with most 
influence on the transport of the granular medium is the Coulomb coefficient,
i.e. an increase of the restitution coefficient from $e=0.5$ to $e=0.99$
leaves the pressure drop unchanged.
\begin{figure}[h]
\begin{center}
\includegraphics[width=10cm]{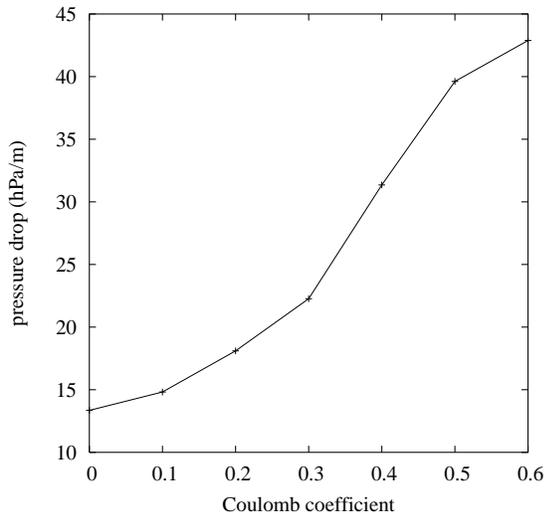}
\end{center}
\caption{\label{fig:cc-dp}%
Dependence of the pressure drop on the Coulomb coefficient,
the superficial gas velocity is $1\,m/s$.
For high Coulomb coefficient ($\mu>0.5$)
a transition from plug conveying to bulk transport can be observed.
}
\end{figure}
As one can see in figure~\ref{fig:cc-dp} the pressure drop increases with
raising Coulomb coefficients.
For low coefficients, $ 0<\mu\le 0.5$, plug conveying is observed,
for higher Coulomb coefficients the plugs get sticky and coalesces
into a big, slowly moving plug. These plugs are similar to the ones
for low superficial gas velocities (fig.~\ref{fig:rho-W0134}).

\subsection{Plug statistics}

The spatio-temporal image of the porosity along the vertical tube
(fig.~\ref{fig:rho-W0126}) provides a rough picture of plugs and their
movement along the tube. A statistical approach is necessary to get
more precise values.
Properties of interest are the porosity and the granular velocity within a plug,
the plug length, and their dependence on the vertical position $z$ of the plug
within the tube.
To get some average values for the porosity and the granular velocity,
the tube was segmented into horizontal slices of height $3.5\,mm$.
For each slice the average porosity and granular velocity has been computed
every $0.01\,s$.
The contribution of a particle has been weighed by the volume occupied
by that particle within a given slice.

The resulting vertical porosity has been used to identify plugs.
Every region with a porosity lower than 0.6 is defined to belong to a plug.
\begin{figure}[h]
\begin{center}
\includegraphics[width=10cm]{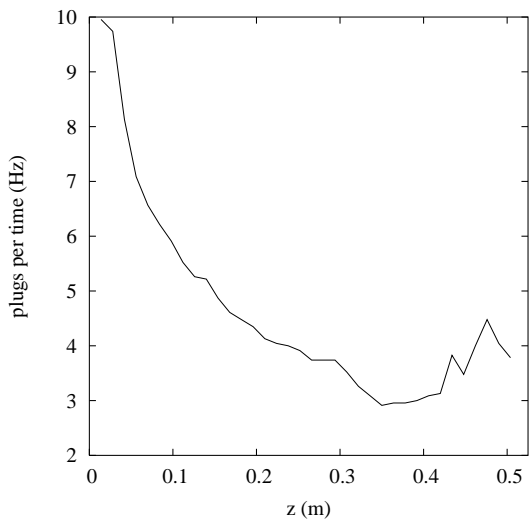}
\end{center}
\caption{\label{fig:W0126_rate_his}%
Number of plugs per time as a function of height $z$.
The corresponding spatio-temporal image is shown in figure \ref{fig:rho-W0126}.
The total tube height is $52.5\,cm$, default parameters are used.
The data is averaged over $23\,s$.
}
\end{figure}

Figure \ref{fig:W0126_rate_his} shows the number of plugs per time
as a function of height.
At the bottom of the tube the incoming granular medium forms into a lot
of small plugs. Even though their velocity is low, the resulting
plug rate is high.
As can be seen in figure~\ref{fig:rho-W0126}, plugs collide
along the tube and merge into
bigger plugs. The merging of plugs reduces the plug rate
along the tube. The smooth decrease of the plug rate indicates that
there is no preferred height for the collision and merging of plugs. 

For each plug the center of mass, the minimal porosity,
the maximal granular velocity and the plug length have been computed.

\begin{figure}[h]
\begin{center}
\includegraphics[width=10cm]{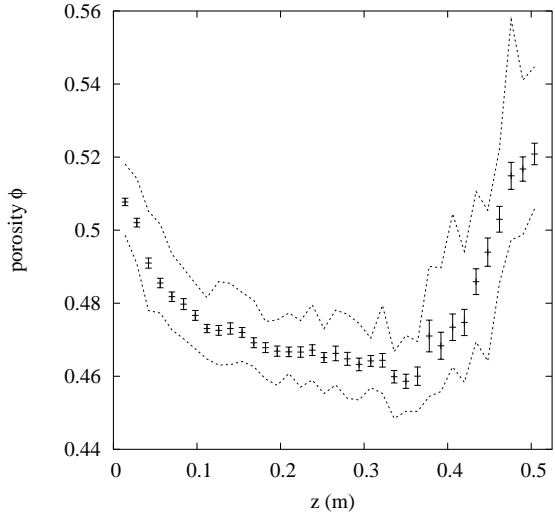}
\end{center}
\caption{\label{fig:W0126_rho_his}%
Minimal porosity found in plugs at height $z$,
corresponding to the plug rate shown in figure \ref{fig:W0126_rate_his}.
The bars denote the uncertainty of the average, and the dotted lines indicate
the width of the distribution;
at each value of $z$, half of the observed plugs have a porosity 
lying between the upper and the lower dotted lines.
The plug porosity decreases at the bottom when the granulate enters the tube
($z<0.1\,m$) and increases at the top of the tube ($z>0.4\,m$),
when the preceding plug has been removed.
}
\end{figure}

Figure \ref{fig:W0126_rho_his} shows 
the minimum plug porosity as a function of the vertical position $z$ of a plug. 
At each value of $z$, the mean porosity and its uncertainty 
(standard deviation divided by the square root of the number of plugs)
were calculated.
These quantities are shown by the bars in Figure 10.
To show the distribution of porosity about the mean,
the two dotted lines were added.
At each height $z$, half of the plugs have a porosity lying between 
these two lines. 
The same analysis was carried out for the data in Figures 11 and 12.
As one can see in figure~\ref{fig:W0126_rho_his} at the left of the graph,
the granular medium is inserted at the bottom
of the tube with a porosity of 0.51.
From there the porosity decreases quickly to about 0.47 at a height $z=0.1\,m$
and then remains almost constant until $z=0.3\,m$.
At the top of the tube ($z>0.3\,m$) the porosity increases until the
grains leave the simulation space at $z=0.525\,m$.

\begin{figure}[h]
\begin{center}
\includegraphics[width=10cm]{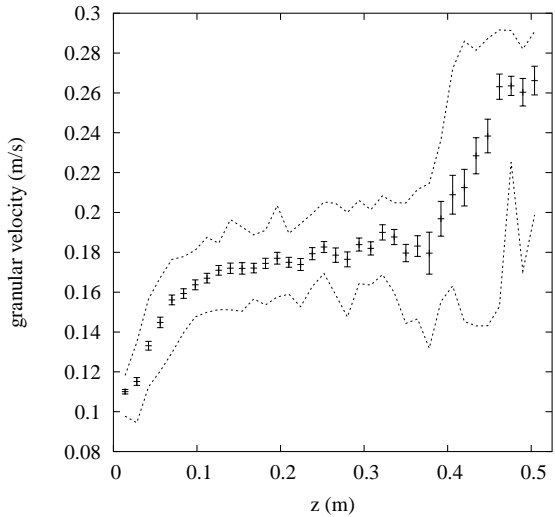}
\end{center}
\caption{\label{fig:W0126_uxx_his}%
Maximal granular velocity found in plugs at height $z$,
corresponding to the plug rate shown in figure \ref{fig:W0126_rate_his}.
The bars and dotted lines have the same meaning as in 
figure~\ref{fig:W0126_rho_his}.
The granular velocity saturates in the middle of the tube, the increase
of the granular velocity at the bottom and at the top of the tube
is due to the boundary effects described in figure \ref{fig:W0126_rho_his}.
}
\end{figure}

Figure \ref{fig:W0126_uxx_his} shows the corresponding particle velocity.
The change in porosity comes along with an increase of the granular velocity
within the plugs.
The granulate is inserted at the bottom with an initial velocity of $0.04\,m/s$.
The granular velocity saturates to a final
velocity for the granular velocity at a height of $0.1\,m$.
At a height of about $0.35\,m$ the granular medium accelerates
until the grains leave the tube.

An explanation for the final constant plug velocity can be derived
using the balance equation for the forces on a plug:
\begin{equation}
 F=-mg-F_c+\alpha(\phi)(v_g-u).
\end{equation}
where $m$ is the mass of a plug, $g$ is the gravity constant,
$F_c$ is the force on the plug through the friction with the wall and
the collisions of particles falling onto the plug. The last term 
$\alpha(\phi)(v_g-u)$ is the drag force on the plug, which is proportional
to the relative velocity between the particles and the gas 
within the plug.
At the middle of the tube the porosity of the plug $\phi$ is constant.
On small time scales the mass $m$ and
the gas velocity $v_g$ can be assumed constant.
Only the drag force depends on the granular velocity $u$ through
d'Arcy's law.
For a certain granular velocity the drag force balances
the sum of gravitational and friction forces. If there is
no acceleration of the plug, then $u$ remains constant.
The solution is stable under small fluctuations in $u$.

\begin{figure}[h]
\begin{center}
\includegraphics[width=10cm]{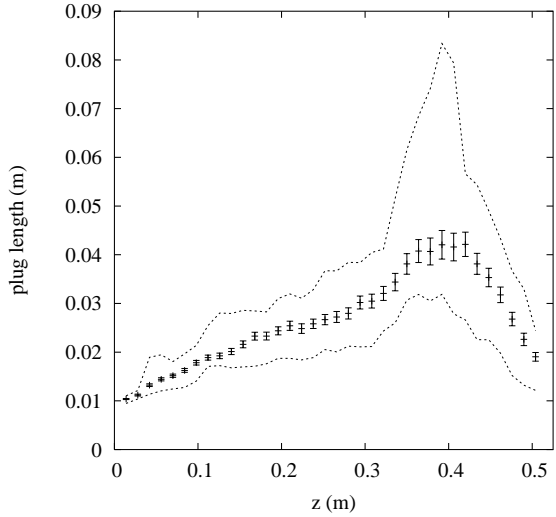}
\end{center}
\caption{\label{fig:W0126_l_his}%
Mean plug length at tube height $z$,
corresponding to the plug rate shown in figure \ref{fig:W0126_rate_his}.
The bars and dotted lines have the same meaning as in 
figure~\ref{fig:W0126_rho_his}.
The average plug length increases along the tube, consistent with the
decrease in the plug rate (fig.~\ref{fig:W0126_rate_his}).
}
\end{figure}
Figure \ref{fig:W0126_l_his} shows the mean plug length along the tube.
The average plug length is increasing along the tube, due to merging
of plugs. Unlike the previous cases no equilibrium length is found.
The increase of the plug length combined with
the decrease of the number of plugs per time (fig.~\ref{fig:W0126_rate_his})
conserves the mass flux of the granulate. 

As can be seen in figure~\ref{fig:rho-W0126}, the change
in porosity, velocity and plug length at the top of the tube
can be explained by the lack of grains falling 
onto the top of the uppermost plug.

The diagrams in figures~\ref{fig:W0126_rho_his} to \ref{fig:W0126_l_his} imply, that
there is a typical porosity, granular velocity and length
of plugs for a given height and a characteristic plug profile exists.
In the following averaged vertical and radial profiles of plugs
at the height of $0.26\,m$ are discussed.
To get some sensible profiles the plugs were selected
with length and granular velocity close to the mean values
($L_t=0.024\pm0.01\,m$, $u=0.16\pm0.02\,m/s$).

\clearpage
\subsection{Vertical plug profiles}
While in experiments recognizing and measuring parameters 
for global plug conveying are rather simple,
the measurement of profiles for individual plugs remains
a nearly impossible task. So one of the reasons for simulating plug conveying
is to provide a detailed picture of what happens within a plug, and how
parameters like porosity, granular velocity or shear stress change along
the plug.

\begin{figure}[h]
\begin{center}
\includegraphics[width=10cm]{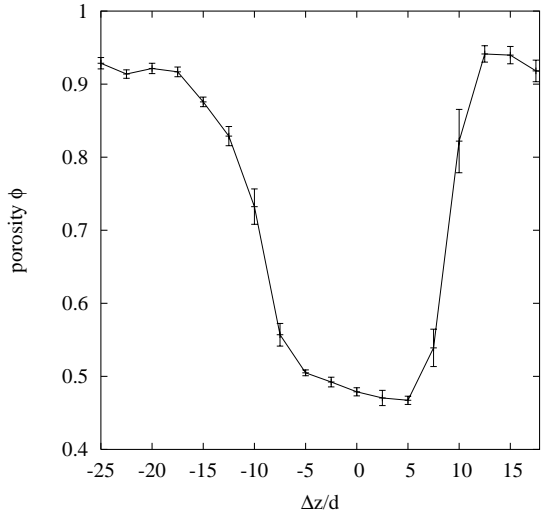}
\end{center}
\caption{\label{fig:rho_x}%
Vertical porosity profile along an averaged plug, containing about 320 particles,
positioned at the middle of the tube. The profile was averaged over 
seven plugs with
granular velocity $0.16\pm0.02\,m/s$ and plug length $0.024\pm0.01\,m$.
The data belong to the simulation displayed in figure~\ref{fig:rho-W0126}.
The horizontal axis denotes the relative vertical position $\Delta z$ 
along the tube with respect to the center of mass. The vertical position
is given in multiples of the particle diameter $d=1.4\,mm$.
}
\end{figure}

\begin{figure}[h]
\begin{center}
\includegraphics[width=10cm]{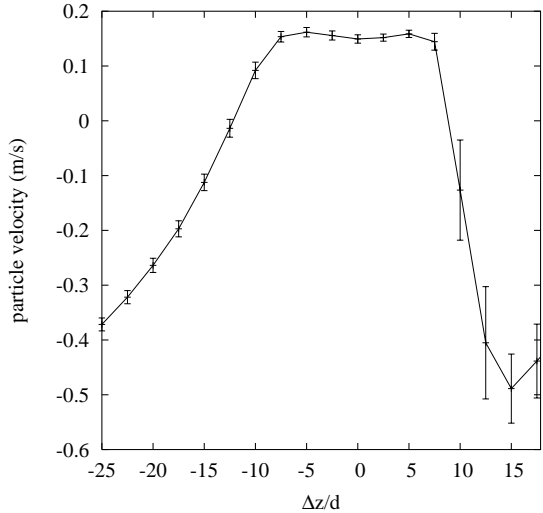}
\end{center}
\caption{\label{fig:uxx_x}%
Velocity of the granular medium along 
the averaged plug of figure~\ref{fig:rho_x}.
Inside the plug ($|\Delta z/d|>7$) the velocity is constant,
outside the granular medium accelerates downwards.
}
\end{figure}

\begin{figure}[h]
\begin{center}
\includegraphics[width=10cm]{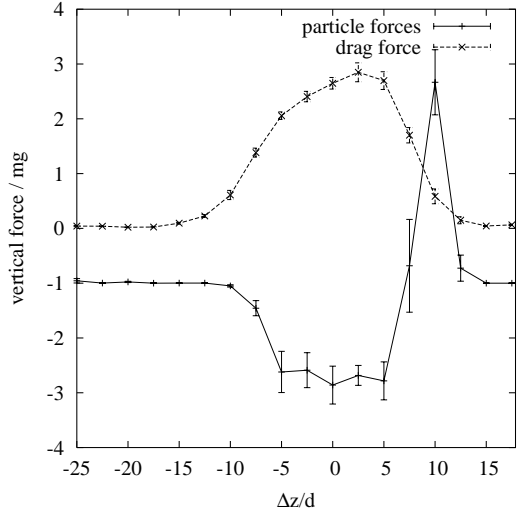}
\end{center}
\caption{\label{fig:force_x}%
Forces acting on particles along the averaged plug 
in figure~\ref{fig:rho_x}.
Interparticle forces, gravity and friction are summed up and displayed
as particle force. The drag force corresponds to the pressure drop of the gas.
Outside of the plug ($|\Delta z/d|>15$) the particle force is $mg$,
indicating that the particle are freely falling.
The peak in the particle force corresponds
to the collision region, where particles fall onto the plug.
}
\end{figure}

\begin{figure}[h]
\begin{center}
\includegraphics[width=10cm]{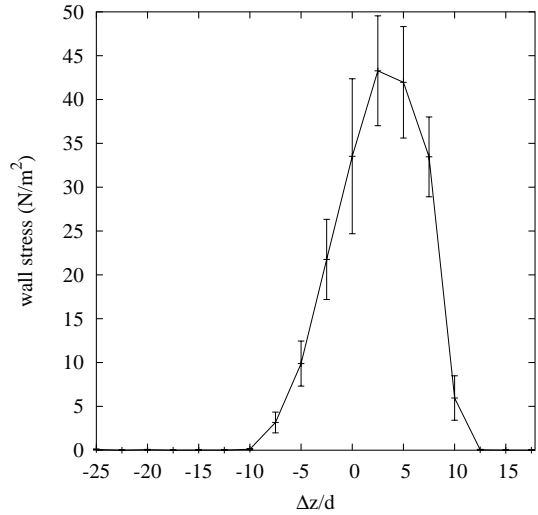}
\end{center}
\caption{\label{fig:wallstress_x}
Stress between the wall and the granular medium along the averaged plug
in figure~\ref{fig:rho_x}.
The wall stress is limited to the plug, where the porosity is low ($<70\%$).
}
\end{figure}

\begin{figure}[h]
\begin{center}
\includegraphics[width=10cm]{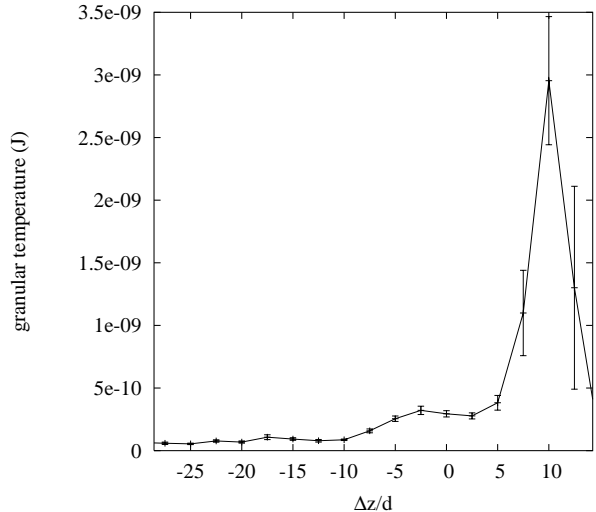}
\end{center}
\caption{\label{fig:eneryyz_x}%
Granular temperature along the averaged plug in figure~\ref{fig:rho_x}.
The granular temperature is only non-zero near the collision region 
between the plug and the falling particles at the top of the plug.
Due to the large damping ($e=0.5$) the granular temperature is dissipated
quickly and remains on a low level within the plug ($|\Delta z/d|> 7$).
}
\end{figure}

A porosity profile of an averaged plug is shown in Figure \ref{fig:rho_x}.
The provided profiles (fig.~\ref{fig:rho_x}-\ref{fig:uxx_r}) 
average over seven plugs.
These plugs were taken from the middle of the tube $z=0.26\,m$ with
granular velocity $0.16\pm0.02\,m/s$ and plug length $0.024\pm0.01\,m$.
The coordinate $\Delta z$ denotes the relative vertical position along the tube 
with respect to the center of mass.
At the top of the plug, on the right hand side of figure \ref{fig:rho_x},
the porosity decreases quickly from 90\% to $45\%$.
In the middle the porosity of the plug remains almost constant,
and increases slowly at the bottom.
The minimum porosity is bigger than a random dense packing 
due to near wall effects.

Figure \ref{fig:uxx_x} displays the velocity profile of
the granular medium in the region of the plug shown
in Figure \ref{fig:rho_x}. One can distinguish four different
regions: Above the plug ($15\le \Delta z/d$)
the granulate particles are falling downwards, the porosity is high.
These particles originate from the bottom of an upper plug.
The drag force cannot support their weight due to the high porosity.
At the top of the plug the particles collide with the low porosity region 
and decelerate rapidly ($7<\Delta z/d\le15$).
The granular medium gets denser as particles are added to the plug.
In the following, this region at the top of the plug 
will be called collision region.
Inside the plug, where the porosity settles to a low value, the granulate
velocity is almost constant ($|\Delta z/d|>7$).
At the bottom ($-10<\Delta z/d\le-7$) of the plug the granular medium starts 
to accelerate downwards again.
Thus the plug is always loosing material at the bottom.
This region will be called disintegration region.

The trajectory of a single particle through the plug can be sketched
by a snapshot of the vertical forces within a plug. Figure \ref{fig:force_x}
displays the drag force $F_d$ and the sum over the interparticle,
gravity and friction forces on a particle (here called particle forces) $F_p$.
These forces are averaged over the horizontal plane.
Above the plug, the particle accelerates constantly due to gravity.
The drag force on a single particle can be neglected, as
the particle velocity is far from the terminal velocity.
The high peak in the particle force marks the collision region.
The huge fluctuations in the particle force arise, 
when the falling particle collides with the top of the plug front.
Within the collision region the porosity decreases and 
therefore the drag force increases.
Below the collision region, within the plug,
the drag force and the particle force balance each other.
At the bottom of the plug
the friction force drops to zero and gravity and drag force
are in equilibrium. Below this point the disintegration region begins,
the granular velocity within the plug decreases, the porosity increases.
After a short distance the drag force is zero and the particles are falling
freely again.

The normal wall stress corresponding to the plug in Figure \ref{fig:rho_x}
is shown in Figure \ref{fig:wallstress_x}. Due to the implementation
of the simulation and the high damping the normal wall stress is proportional
to the vertical shear stress (Coulomb friction).
The wall stress is limited to the plug including the collision 
and disintegration region. The wall stress increases from zero
within the collision region and then slowly decreases, reaching zero
in the disintegration region.
The maximal wall stress is of the same order of magnitude 
as the pressure imposed on the plug front by the falling granulate 
$P=\rho_s(1-\phi)v^2=937\,kg/m^3\cdot0.07\cdot(0.7\,m/s)^2 \approx 32Pa$.

Figure \ref{fig:eneryyz_x} shows the granular temperature along a plug.
The granular temperature is the average kinetic energy of particles
minus the kinetic energy of the motion of their center of mass.
Appreciable granular temperatures occur in the collision region.
Due to the high damping ($e=0.5$) these temperatures are confined to a small
region at the top front of the plug. There is a finite granular temperature 
within the plug, the temperature drops nearly to zero at the bottom of the plug.

\subsection{Radial plug profiles}
Another point of view on the plug is given by the radial profiles.
The volume used for averaging at a given radius
is a cylindrical ring with a height of
1.5 particle diameters and a width of 0.25 particle diameters.
The porosity at a given range is defined to be the volume fraction
of the gas within the cylindrical ring.

\begin{figure}[h]
\begin{center}
\includegraphics[width=10cm]{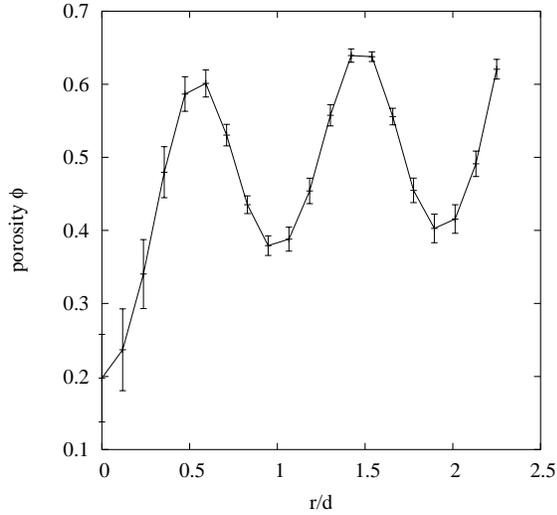}
\end{center}
\caption{\label{fig:rho_r}%
Radial porosity profile through an averaged plug 
at the height $z=0.26\,m$. The same plugs were used for averaging
as for the vertical profiles in figure~\ref{fig:rho_x}.
The radial distance $r$ from the middle of the tube is given
in units of the particle diameter $d=1.4\,mm$.
The porosity minima correspond to three layers of particles moving
upwards, parallel to the tube wall.
}
\end{figure}
As a result of this high radial resolution it is possible to resolve
layers of granular particles. 
As can be seen in figure~\ref{fig:rho_r},
the plugs, averaged for the vertical profiles, are highly ordered
in the radial direction. The radial profiles are taken from the middle of the
averaged plugs, at half height of the tube ($z=0.26\,m$).
The porosity minima in figure~\ref{fig:rho_r} correspond to 
three layers of particles.
This agrees well with the ratio between the tube and particle diameter, 
which implies that five particles fit into the tube horizontally.
The minimum at $r/d=0$ corresponds to a vertical
string of particles in the middle of the tube, the layer at $r/d=2$ touches
the wall of the tube.
For higher radii $r$ the porosity increases to one.
The large error at the center of the tube is due to the smaller volume involved
for averaging.

\begin{figure}[h]
\begin{center}
\includegraphics[width=10cm]{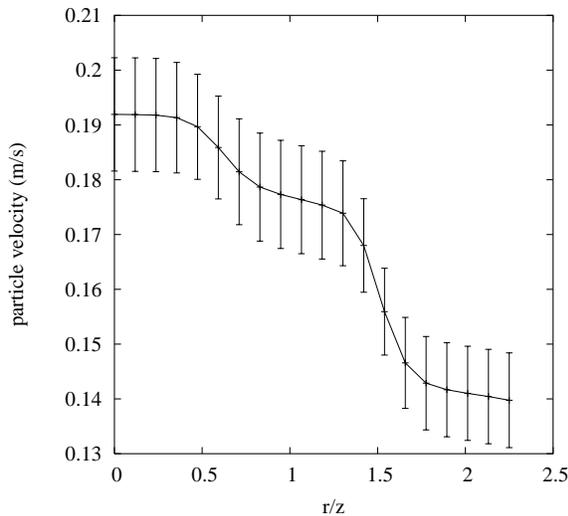}
\end{center}
\caption{\label{fig:uxx_r}%
Radial profile of the granular velocity for an averaged plug 
in figure~\ref{fig:rho_r}.
The profile shows three layers of particles
moving upwards with different velocities. The layer along the wall is
the slowest due to the friction with the wall.
}
\end{figure}

The radial profile of the granular velocity (fig.~\ref{fig:uxx_r}) shows
that the layers observed in the porosity profile move with different
velocities. The layer at the wall, which has the largest impact
on the total averaged granular velocity, moves slowest,
due to the friction with the wall.
\clearpage
%
%
%
%
\section{Conclusion}
In this paper a simple model~\cite{MCN0003} with coupled grain and gas flow
has been applied to pneumatic transport.
The implementation is three-dimensional; rotation
and Coulomb friction are taken into account.
The flux of gas and grains is set by the boundary conditions.
Plug conveying is observed. 
The simulation used for plug statistics and profiles contained
on average 3600 particles in a tube of length $0.525\,m$ 
and diameter $7\,mm$.
The simulation lasted $23\,s$;
during this time 90 different plugs were observed
at the middle of tube, each with about 300 particles.
Additionally 24 simulations were preformed to obtain the characteristic curves.

The behavior of the granular medium and the gas is consistent with 
experimental observations.
The pressure drop, mass flux and superficial velocity can be obtained in
quantitative agreement with experiment 
by introducing an effective viscosity $\eta$ and an effective friction $\mu$. 
The effective viscosity reflects the increased momentum transport
in the gas due to the turbulent flow around the grains.
The effective friction reflects the complex interplay between sliding,
rolling and static friction.
By using the model,
large numbers of plugs could be studied, and their porosity, velocity,
and size were measured as functions of height.

The simulation results imply that the formation of the plugs
at the bottom of the tube occurs spontaneously.
There is a well defined preferred velocity of the plugs,
which is independent of the plug size and the tube length.
The average plug length increases along the tube due to the merging
of plugs.
To confirm experimentally the constant velocity 
and the increase of plug length,
statistics of plug length and velocity at different heights along the tube are needed.
The results also show that boundary effects at the bottom and the top
of the tube have to be taken into account. 
For the experimental parameters the influence of the boundaries is limited 
to a distance of $0.1\,m$,
which is about the distance between two plugs in the middle of the tube. 
Plugs  accelerate in these boundary regions.
The acceleration at the upper boundary arises due to the lack of
grains falling on the uppermost plug.
This implies that the momentum transfer due to falling grains can not be neglected.
A model of plug conveying must take into account that
these falling grains reduce the plug velocity.

Also a detailed view into an average plug was presented.
In contrast to experimental setups were are not limited to measure the 
plug properties only at few locations.
Experimental results usually provide the plug properties as a function of
time at a given position along the tube, with the disadvantage that the
plug profile is distorted by the relative motion of granulate along
the plug. 
The vertical profiles were given for porosity, granular velocity,
interparticle and drag forces, wall stress and granular temperature.
In the experiment, these parameters usually are not accessible along the whole tube.

Additionally the radial profile is given at high resolution.
The results show that the grains are ordered in radial direction.
The grains arrange themselves in layers along the tube wall.
These layers have different velocities (up to 40\% difference). 
Experimental measurements of the granular velocity using a CCD-Camera
must take into account that they measure the velocity of the outer, slowest
layer.

In conclusion, the model presented here is a useful tool for investigating
plug conveying.
It is fast and flexible enough to make parameter studies 
on full featured plug conveying
and provides at the same time access to the plug properties at the level
of grains.
It can obtain both characteristic curves and plug profiles.
Note also that the model can be easily used to describe horizontal transport,
simply by rotating the gravitational acceleration.
Future optimizations and faster computers 
will permit this model to be applied to
industrial-sized systems.
%
%
%
%
\section{Acknowledgments}
We thank Ludvig Vinningland and Eirik Flekk{\o}y for much help in the
fluid solver, Thomas Ihle and Adriano de Oliveira Sousa 
for their general help and advice.
This research was supported by DFG (German Research Community)
contract HE 2732/2-1 and HE 2732/2-3.
%
%


\end{document}